# High temperature strength retention of Cu/Nb nanolaminates through dynamic strain ageing

Zhilin Liu[a, b], J. Snel[b], T. Boll[c], J.-Y. Wang[b, d], M. A. Monclús[b], J. M. Molina-Aldareguía[b], J. LLorca[b, d, *]

[a]Light Alloy Research Institute, College of Mechanical and Electrical Engineering, Central South University, 410083, P.R. China

[b]IMDEA Materials Institute, C/Eric Kandel 2, 28906, Getafe, Madrid, Spain

[c]Institute for Applied Materials IAM-WK and Karlsruhe Nano Micro Facility, Karlsruhe Institute of Technology, 76344 Eggenstein-Leopoldshafen, Germany

[d]Department of Materials Science, Polytechnic University of Madrid/Universidad Politécnica de Madrid, E.T.S. de Ingenieros de Caminos, 28040 Madrid, Spain

Corresponding author: javier.llorca@imdea.org

**Abstract**

The mechanical properties of Cu/Nb metallic nanolaminates with different layer thickness (7, 16, 34 and 63 nm) were studied by means of micropillar compression tests from room temperature to 400 ºC. Both strain-rate jump and constant strain rate tests were carried out and they showed evidence of dynamic strain ageing in the nanolaminates with 7, 16 and 34 nm layer thickness deformed at 200 ºC. Dynamic strain ageing was accompanied by a reduction of the strain rate sensitivity to 0, high strength retention at 200 ºC and the development of shear localization of the deformation at low strains (5%-6%) that took place along the Nb layers in the nanolaminates. Atom probe tomography of the deformed specimens revealed the presence of O in solid solution in the Nb layers but not in the Cu layers. Thus, diffusion of O atoms to the mobile dislocations in Nb was found to be the origin of the dynamic strain ageing in the Cu/Nb nanolaminates around 200 ºC. This mechanism was not found at higher temperatures (400 ºC) because deformation was mainly controlled by stress-assisted diffusion in the Cu layers. This discovery shows a novel strategy to enhance the strength retention at high temperature of metallic nanolaminates through dynamic strain ageing of one the phases.

# 1. Introduction

Metallic nanolaminates (MNL) are novel materials formed by alternating layers of dissimilar metals, with a thickness in the range of nm. Their dominant deformation and fracture mechanisms are dramatically changed - as compared with conventional metallic materials - due to the high density of interfaces, leading to superior properties for structural applications, such as very high flow strength (> 1 GPa), high indentation hardness, excellent ductility, good radiation damage resistance and thermal stability, as well as promising fatigue/failure resistance [1-6]. Nevertheless, the actual use of MNLs in structural applications in engineering was conditioned by the processing route (magnetron sputtering) that limited the thickness and area of nanolaminates. This limitation was recently overcome with the successful manufacturing of MNL through accumulated roll bonding (ARB), a traditional metallurgical route that allows the production of laminates with large area and thicknesses in the range of mm [7-9].

The mechanical properties and the associated deformation mechanisms of MNLs at room temperature have been extensively studied by means of nanomechanical testing techniques [1, 6-14], particularly in Cu/Nb MNLs. Nevertheless, the experimental data available on their mechanical properties and strain rate sensitivity at high temperature are very scarce [15-17]. Snel *et al.* [17] characterized the deformation mechanisms of ARB Cu/Nb MNLs at 25 ºC and 400 ºC and found that confined layer slip was the dominant mechanisms in this temperature range in so far the layer thickness was above a critical value (20-35 nm) which depended on the strain rate. Below this thickness, the deformation shifted from dislocation transmission at the interfaces at 25 ºC to dislocation climb at 400 ºC and this change was associated with a large reduction in strength with temperature, which did not occur in the case of confined layer slip. As a result, an optimum layer thickness of 20-35 nm was found to provide an outstanding strength, in the range 0.8 to 1.2 GPa, at 400 ºC while the layered microstructure remained stable up to 700 ºC [18].

These results reinforce the potential of ARB MNLs for structural applications and support the exploration of other strengthening mechanisms that can operate at high temperature. In particular, strength of Nb is known to increase in the presence of O in the temperature range ≈ 150 – 300 ºC due to dynamic strain aging (DSA) [19-21]. DSA arises from the dynamic interaction of solute atoms and mobile dislocations. Diffusion of solute atoms to the dislocations may occur during the waiting time of the mobile dislocations to overcome the obstacles for a given temperature (that controls the diffusion of the solute atoms) and strain rate (that determines the average dislocation mobility). Thus, the stress to keep the dislocations moving will increase due to DSA at constant strain rate and the relation between stress, strain and strain rate will be affected [22]. In general, DSA leads to an increase in the flow stress together with serrated flow behaviour and to a reduction in the strain rate sensitivity (which may become negative) favouring strain localization [23-25].

In this investigation, evidence of DSA is provided for the first time in MNLs. The influence of DSA due to the presence of O in the Nb layers is analyzed between 25 ºC and 400 ºC for strain rates in the range of $10^{-2}$/s to $10^{-4}$/s in Cu/Nb MNLs processed by ARB with average layer thicknesses between 7 and 63 nm. DSA was clearly observed in the MNL at 200 ºC, leading to a significant strength retention at this temperature. The relationship between the microstructure of the MNLs with the deformation mechanisms was clearly established, opening the path to new strategies to develop MNLs with improved mechanical properties at high temperature.

# 2. Experimental Section

2.1. Preparation of micropillars and TEM samples

Cu/Nb bulk MNLs with average layer thickness of 7, 16, 34 and 64 nm were manufactured by ARB from two polycrystalline sheets of Cu (99.99 wt.%) of 0.5 mm in thickness and one polycrystalline sheet of Nb (99.94 wt.%) of 1 mm in thickness. After a surface cleaning treatment, the Cu and Nb sheets were stacked in a sandwich-like structure for the first roll bonding at room temperature, leading to a thickness reduction of 60%. Then the sample was halved, cleaned and restacked, whereupon the ARB process continued by repeating these steps. When the individual layer thickness reached 200 nm, the samples were only subjected to conventional rolling (no cutting and restacking steps) and the thickness reduction per rolling step was 10%. More details of the forming process can be found elsewhere [7]. While the layer thickness is fairly constant in MNL manufactured by magnetron sputtering, this is not the case in ARB MNLs which show a wide distribution of layer thicknesses around the average value. Samples for mechanical characterization were cut from the bulk, followed by mechanical grinding and polishing. The final mechanical polishing process was carried out in neutral diamond suspensions (40 nm particle size). Cu/Nb micropillars were milled out from as-polished bulk samples by means of a dual-beam scanning electron microscope (FEI Helios NanoLab 600i). Ga$^+$ ions were used for milling. Micropillars with either square or circular cross-section were prepared by annular milling according to an in-house milling script. FIB milling was carried out in successive steps with decreasing currents of 21 nA, 2.5 nA, 0.24 nA, 80 pA and 40 pA to improve the surface finish and reduce the ion implantation. The lateral dimensions (either diameter or side length) were in the range 4.6 µm to 5.3 µm, with an aspect ratio of 2.2 – 2.8 and a taper angle of 0.2º –1.6º. TEM samples were milled and extracted from the deformed micropillars following the procedure detailed in Figs. 1a-1d. Post-thinning of the TEM samples was carried out at an acceleration voltage of 5 kV to improve their quality.

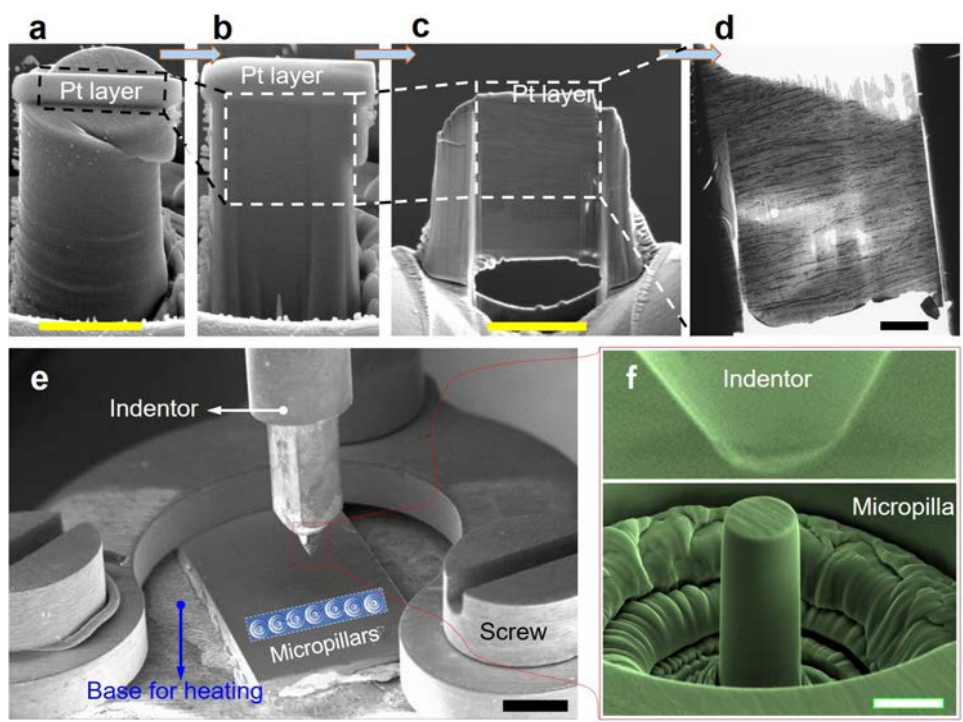

**Fig. 1.** (a, b, c) Focused-ion beam milling of a thin foil parallel to the micropillar containing the shear band was lifted from the as-deformed micropillars. (d) Transmission electron microscopy micrograph of a representative thin foil. (e, f) Schematic illustration of in-situ microcompression tests. The scale bars in (a) is 1 mm, in (b) 5 µm, in (c) 4 µm, in (e) 5 µm, and in (f) 1 µm.

2.2. In-situ microcompression tests

Compression tests of micropillars were performed with a 10-µm diamond flat punch using a Hysitron® PI87HT nanoindenter combined with the TriboScan™ software. The tests were carried out within a scanning electron microscope (SEM; Zeiss EVO MA15). The test temperature was adjusted using a LakeShore® 336 temperature controller. The sample was placed on the heating plate using a U-shaped clamp in combination with carbon-based paint to maintain optimum contact.

Experimental setup is shown in Figs. 1e and 1f. The sample was kept for several minutes at the target test temperature (either RT, 200 ºC or 400 ºC) to stabilize the temperature and minimize the thermal drift before mechanical deformation was applied. Strain-rate jump tests were performed in Cu/Nb MNL micropillars with layer thickness of 7, 16, 34 and 64 nm. Strain-rate jump tests were selected because the strain rate sensitivity can be determined from a single compression experiment by introducing several strain jumps and this strategy minimizes the inherent scatter associated with nanomechanical testing. Moreover, the reduction in testing time eliminates problems associated with thermal drift. Displacement rates were controlled at 1.5 nm/s, 15 nm/s, 150 nm/s, which contributed to an equivalent strain rate of $10^{-4}$/s, $10^{-3}$/s and $10^{-2}$/s, as indicated in Ref. [17]. Evidence of DSA was found at 200 ºC and strain rate of $10^{-3}$/s for MNL with layer thicknesses ≤ 34 nm. Therefore, constant strain-rate experiments ($10^{-3}$/s) were carried out within a scanning electron microscope in circular micropillars of Cu/Nb MNL with 7 nm layer thickness to obtain more detailed evidence of the development of DSA in the temperature range RT to 400 ºC.

2.3. Site-specified APT sample preparation from shear zones

Atom probe tomography (APT) can be provide a very accurate measurement on the O content within the Cu and Nb layers of the MNL. Using a dual-beam scanning electron microscope (Zeiss Auriga 60), APT needles were prepared from the micropillars of the MNL with 7 nm layer thickness deformed at RT, 200 ºC and 400 ºC using the procedure depicted in Fig. 2. A final cleaning of the microtip was operated at 3 kV to obtain an acceptable level of $Ga^+$ content (< 0.05 at.%).

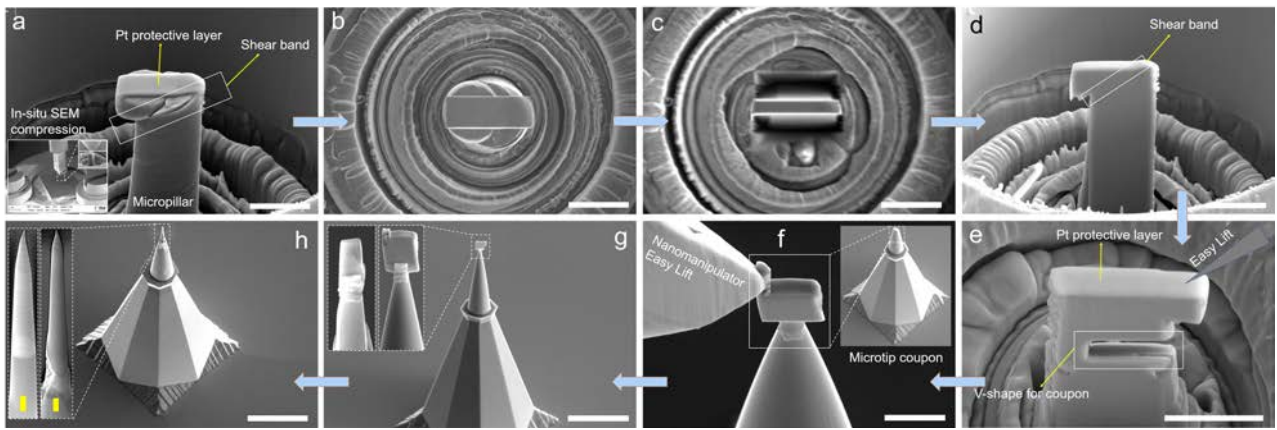

**Fig. 2.** Focused-ion beam (FIB) milling steps to prepare APT needles near the shear bands of Cu/Nb MNL micropillars after the compression tests at RT, 200 ºC and 400 ºC. (a) A protective layer of Pt was deposited on the top surface of the deformed micropillars to avoid ion beam damage. (b, c and d) Milling of undesired regions of micropillars. (e) V-shape milling at the bottom to connect with the APT microtip coupon. (f) Transfer and welding of the microtip coupon using the Easy Lift® nanomanipulator. (g) The V-shape bottom of one shear band was welded onto the microtip coupon. (f) Sharpening of the rough shear bands into a nano-scale needle shape using very small current and voltage with the FIB. The inset of (h) shows two representative APT needles. The white scale bars are 5 μm in (a), (b), (c), (d) and (f), 3 μm in (e), and 40 μm in (g) and (h). The yellow scale bars in the inset of (h) are 500 nm.

2.4. Microstructural characterization by electron microscopy

After in-situ microcompression at different temperatures, site-specified TEM samples were obtained from the shear bands using the dual-beam scanning electron microscope (FEI Helios NanoLab 600i) and the nanomanipulator (Easy Lift™). In order to avoid oxidation, the TEM samples were immediately transferred into transmission electron microscopy (TEM; FEI Talos F200X). Thereafter, microstructures of as-compressed micropillars were analyzed at 200 keV in the TEM using the HAADF-STEM mode as well as EDS.

2.5. Atom probe tomography measurement

APT measurement was operated in a LEAP™ 4000X HR at -213.15 ºC. The pulse fraction and pulse frequency of APT facility were set to be 20% and 200 kHz, respectively. APT data analysis was completed using Cameca software IVAS 3.6.14. The mass spectra analysis was performed from site-specified regions based on the average data of 5 different APT runs.

2.6. Calculation of the mechanical properties

The engineering stress-strain curves were obtained from the corrected load-displacement curves assuming that the pillars were perfect prisms or cylinders of length $L_0$ and cross-sectional area $A_0$ measured as the top cross-sectional area of the pillar. The engineering stress $\sigma_{eng}$ and engineering strain $\varepsilon_{eng}$ were calculated, respectively, as

$$\sigma_{eng} = \frac{P}{A_0} \qquad (1)$$

$$\varepsilon_{eng} = \frac{u_p}{L_0} \qquad (2)$$

where $P$ is the applied force and $u_p$ the corrected pillar displacement, using Sneddon´s correction for instrument and substrate compliance [28].

The strain-rate sensitivity ($m$) can be calculated from the flow stresses ($\sigma_1$ and $\sigma_2$) and the strain rates ($\dot{\varepsilon}_1$ and $\dot{\varepsilon}_2$) before and after each jump according to [17]

$$m = \frac{\ln(\sigma_2/\sigma_1)}{\ln(\dot{\varepsilon}_2/\dot{\varepsilon}_1)}\bigg|_T \qquad (3)$$

The activation volume, $V$, is given by

$$V = \sqrt{3}kT\frac{\partial \ln(\dot{\varepsilon})}{\partial \sigma} \qquad (4)$$

where $k$ is the Boltzmann constant, $T$ the absolute temperature and $\sigma_f$ the flow stress. [26].

# 3. Results

*3.1. Strain-rate jump tests of Cu/Nb MNL*

The mechanical behavior of Cu/Nb MNL with different nanolayer thicknesses was explored at room temperature (RT), 200 ºC and 400 ºC using strain-rate jump tests. The engineering stress- strain curves obtained from micropillar compression tests are plotted in Figs. 1a to 1d for Cu/Nb MNL with average layer thicknesses of 7, 16, 34, and 63 nm, respectively. The initial strain rate was $10^{-3}$/s in all cases. Then, the strain rate in each test was changed systematically following the sequence $10^{-3}$/s → $10^{-4}$/s → $10^{-3}$/s → $10^{-2}$/s → $10^{-3}$/s. Increasing the strain rate led to hardening and decreasing the strain rate to softening in all the Cu/Nb MNLs deformed at RT and 400 ºC. Of course, the effect of the strain rate jumps was more noticeable at 400ºC. However, these jumps in the flow stress were not observed in the stress-strain curves of the Cu/Nb MNL with layer thicknesses of 7, 16 and 34 nm deformed at 200 ºC (Figs. 3a to 3c), indicating that the strain rate sensitivity was negligible in these circumstances. It should be noted, however, that the MNL with 63 nm layer thickness tested at 200 ºC did not show this behavior and the stress-strain curves were similar to those reported at RT and 400 ºC (Fig. 3d).

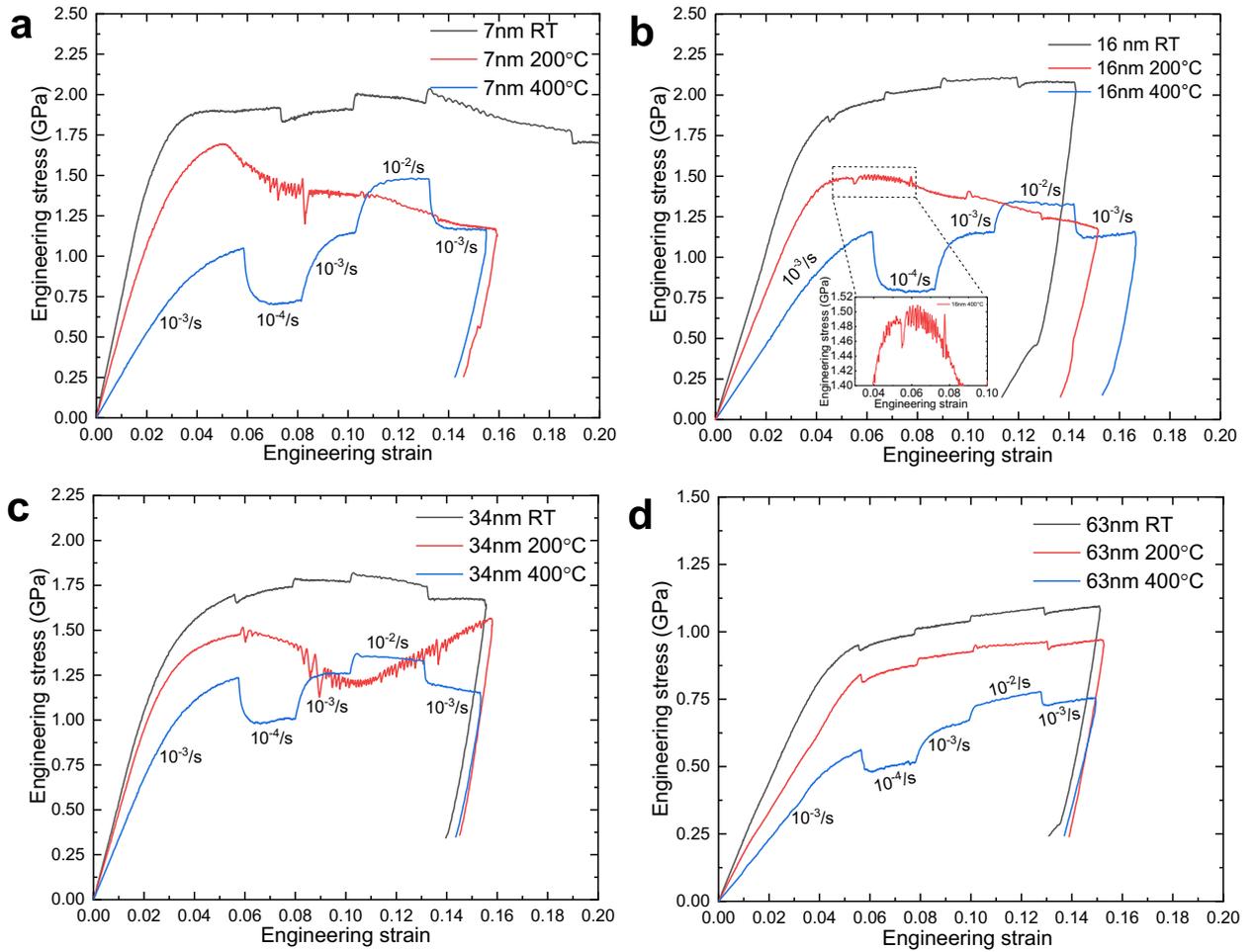

**Fig. 3.** Engineering stress-strain curves of Cu/Nb MNL micropillars with different average layer thickness subjected to strain-rate jump tests deformed at room temperature (RT), 200 ºC and 400 ºC. (a) 7 nm. (b) 16 nm. (c) 34 nm. (d) 63 nm. Unless specified otherwise, the strain-rate jump tests in each figure were loaded in an order of $10^{-3}$/s → $10^{-4}$/s → $10^{-3}$/s → $10^{-2}$/s → $10^{-3}$/s.

The stress-strain curves of the 7, 16 and 34 nm Cu/Nb MNL deformed at 200 ºC presented other similar features. The deformation was homogeneous up to a critical strain of 5%-6%, which also corresponded to a maximum in the flow stress. Afterwards, serrations appeared in the stress-strain curves, which were associated with a reduction in the stress carried by the micropillars. The deformed micropillars were analyzed in the scanning electron microscope and the corresponding micrographs are depicted in Fig. 4. Failure due to shear localization was clearly observed in the Cu/Nb MNL with 7, 16 and 34 nm average layer thickness deformed at RT and 200 ºC but it should be noted that the shear localization developed at low strains (5%-6%) at 200ºC while it required much higher strains at RT (> 12%). Shear localization was not found in the Cu/Nb MNL with 64 nm layer thickness in the whole temperature range nor in the MNL with different layer thicknesses tested at 400 ºC. Moreover, Cu globules were observed on the lateral surfaces of the micropillars deformed at 400 ºC as a result of the activation of stress-assisted diffusion mechanisms [17].

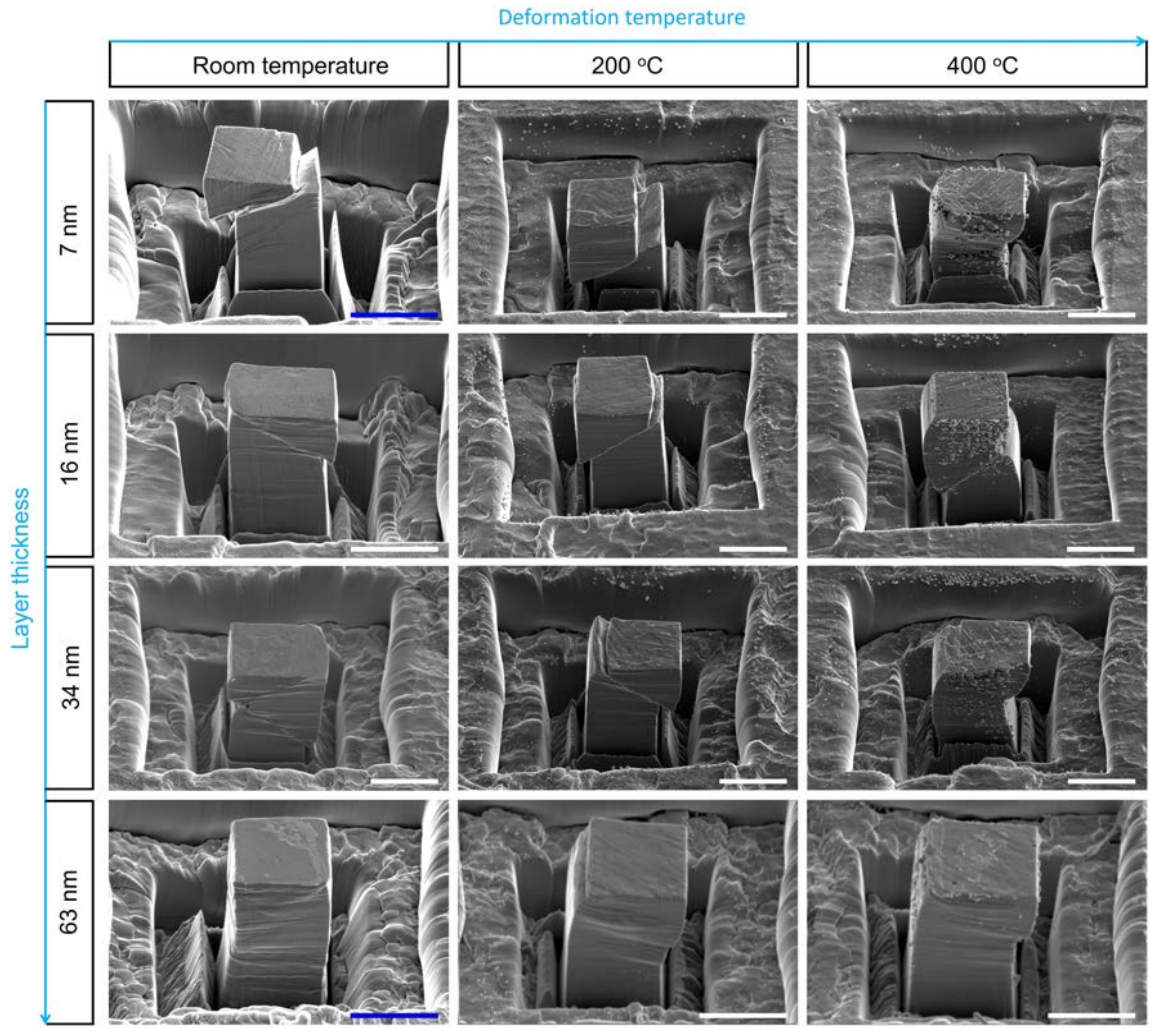

**Fig. 4.** SEM micrographs of the Cu/Nb nanolaminate micropillars with individual layer thicknesses of 7 nm, 16 nm, 34 nm and 63 nm, deformed at room temperature, 200 ºC and 400 ºC, respectively. The last two micrographs for 63 nm are extracted from Ref. [17] for comparison. All scale bars are 5 µm.

The strain-rate sensitivity ($m$) and the activation volume ($V$) were determined from the curves in Fig. 3 following the methodology detailed in the Experimental Section and they are listed in Supplementary Tables 1 and 2. The evolution of the strain rate sensitivity $m$ as a function of the layer thickness and temperature is plotted in Figs. 5a and 5b from the $10^{-3}$/s to $10^{-4}$/s and $10^{-2}$/s to $10^{-3}$/s strain rate jumps, respectively. $m$ was very small (0.001 - 0.006) for all layer thickness at RT and activation volumes were around $20b^3$ ($b$ stands for the Burgers vector) for the MNLs with layer thicknesses of 7 and 16 nm and around $50-70b^3$ when the layer thickness was 34 and 60 nm. The former $V$ were compatible with deformation controlled by dislocation transmission at the interfaces, which takes place at very high stresses (> 1.5 GPa), while the latter $V$ indicated that confined layer slip was the dominant plastic deformation mechanism at RT for Cu/Nb MNL with layer thickness > 30 nm [26]. $m$ decreased to practically 0 at 200 ºC in the MNLs with average layer thickness 7, 16 and 34 nm and increased rapidly afterwards at 400 ºC, reaching strain rate sensitivities close to 0.2 for the MNL with smaller layer thickness. The corresponding activation volumes were small (< $20b^3$), which suggest that deformation was in the power-law creep regime [17]. The strain rate sensitivity of the MNL with 64 nm layer thickness did not decrease at 200 ºC. Instead, it presented a continuous -although limited- increase from RT at 400 ºC. Moreover, the activation volumes for this MNL were compatible with confined layer slip in the whole temperature range ($30b^3$-$100b^3$).

**Table 1** Strain-rate sensitivity *m* and activation volume *V* calculated from compression tests at strain rates between $10^{-4}$ and $10^{-3}$/s. *b* is Burgers vector. Reference data at RT and 400 °C (from [17]) are included for comparison.

| Layer thickness | Strain-rate sensitivity (*m*) | | | Activation volume (*V*) | | |
| --- | --- | --- | --- | --- | --- | --- |
| | RT | 200 °C | 400 °C | RT ($b^3$) | 200 °C ($b^3$) | 400 °C ($b^3$) |
| 7 nm | 0.010±0.001 | ±0 | 0.19±0.01 | 24.5±2.0 | ∞ | 6.2±0.5 |
| 16 nm | 0.010±0.002 | ±0 | 0.17±0.03 | 24.8±4.0 | ∞ | 6.0±0.7 |
| 34 nm | 0.006±0.001 | 0.003 | 0.10 | 50.8±0.8 | 166.0 | 9.7±5.4 |
| 63 nm | 0.007±0.001 | 0.011±0.001 | 0.05±0.02 | 61.6±6.7 | 73.7±2.3 | 28.7±3.9 |

**Table 2** Strain-rate sensitivity *m* and activation volume *V* calculated from compression tests at strain rates between $10^{-3}$ and $10^{-2}$/s. *b* is Burgers vector. Reference data at RT and 400 °C (from [17]) are included for comparison.

| Layer thickness | Strain-rate sensitivity (*m*) | | | Activation volume (V) | | |
| --- | --- | --- | --- | --- | --- | --- |
| | RT | 200 °C | 400 °C | RT ($b^3$) | 200 °C ($b^3$) | 400 °C ($b^3$) |
| 7 nm | 0.015±0.002 | ±0 | 0.10±0.01 | 17.0±1.8 | ∞ | 7.9±0.7 |
| 16 nm | 0.012±0.004 | ±0 | 0.08±0.01 | 22.7±4.9 | ∞ | 11.4±1.5 |
| 34 nm | 0.006±0.001 | 0.003 | 0.05±0.02 | 46.7±6.7 | — | 18.3±5.2 |
| 63 nm | 0.006±0.001 | 0.008±0.001 | 0.026±0.004 | 69.8±11.8 | 100.0±14.6 | 49.0±8.1 |

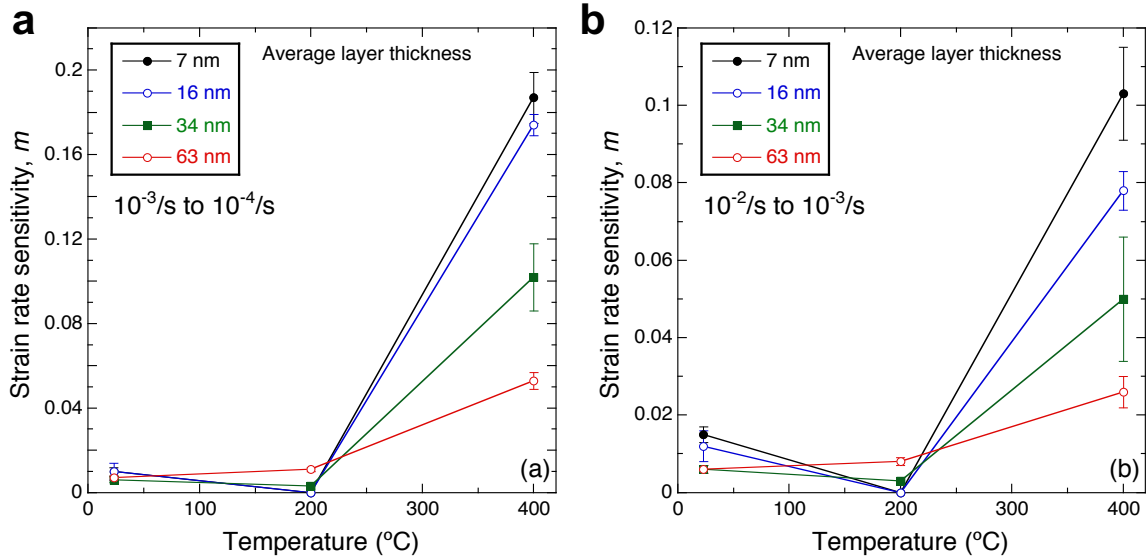

**Fig. 5.** Strain rate sensitivity *m* of Cu/Nb MNL as a function of layer thickness and temperature obtained from strain rate jump tests. (a) Strain rate jump from $10^{-3}$/s to $10^{-4}$/s. (b) Strain rate jump from $10^{-2}$/s to $10^{-3}$/s.

*3.2. Onset of serration of Cu/Nb MNL at constant strain rate*

Cylindrical micropillars of the Cu/Nb MNL with 7 nm layer thickness were compressed within a scanning electron microscope to obtain more direct evidence of DSA. These tests were carried out at constant strain rate of $10^{-3}$/s at RT, 200 °C and 400 °C and the corresponding engineering stress-strain curves are plotted in Fig. 6a. The deformation mechanisms at RT, 200 °C and 400 °C are shown in the Supplementary Videos S1, S2 and S3, respectively, and the final shape of the deformed micropillars is depicted in the scanning electron micrographs in Figs. 6b (RT), 6c (200 °C) and 6d (400 °C). These figures show that final failure of the micropillars took place by the formation of a shear band, but the origin and the mechanisms associated were very different. Deformation was homogeneous up to ≈ 6% strain in the micropillar deformed at RT (Fig. 6a) and the shear localization from the top of the pillar was triggered due to the stress concentration at this location and the limited strain hardening capability of the MNL (see Supplementary Video S1 and Fig. 6b). Nevertheless, shear localization began at much lower strains (≈ 3%) in the micropillars deformed at 200 °C, and it was associated with the development of plastic instabilities in the form of serrations in the stress-strain curve and with the formation of the shear band that was often localized in the middle of the micropillars (see Supplementary Video S2 and Fig. 6c). Finally, the video of the micropillar deformed at 400 °C showed the extrusion of Cu from the lateral surfaces during the last stages of elastic regime, indicating that the dominant deformation mechanisms was stress-assisted diffusion of Cu (Supplementary Video S3). Further deformation led to the appearance of a shear band in the middle of the micropillar, which was associated with the extrusion of Cu. Moreover, globules of Cu were found on the micropillar surface (Fig. 6d).

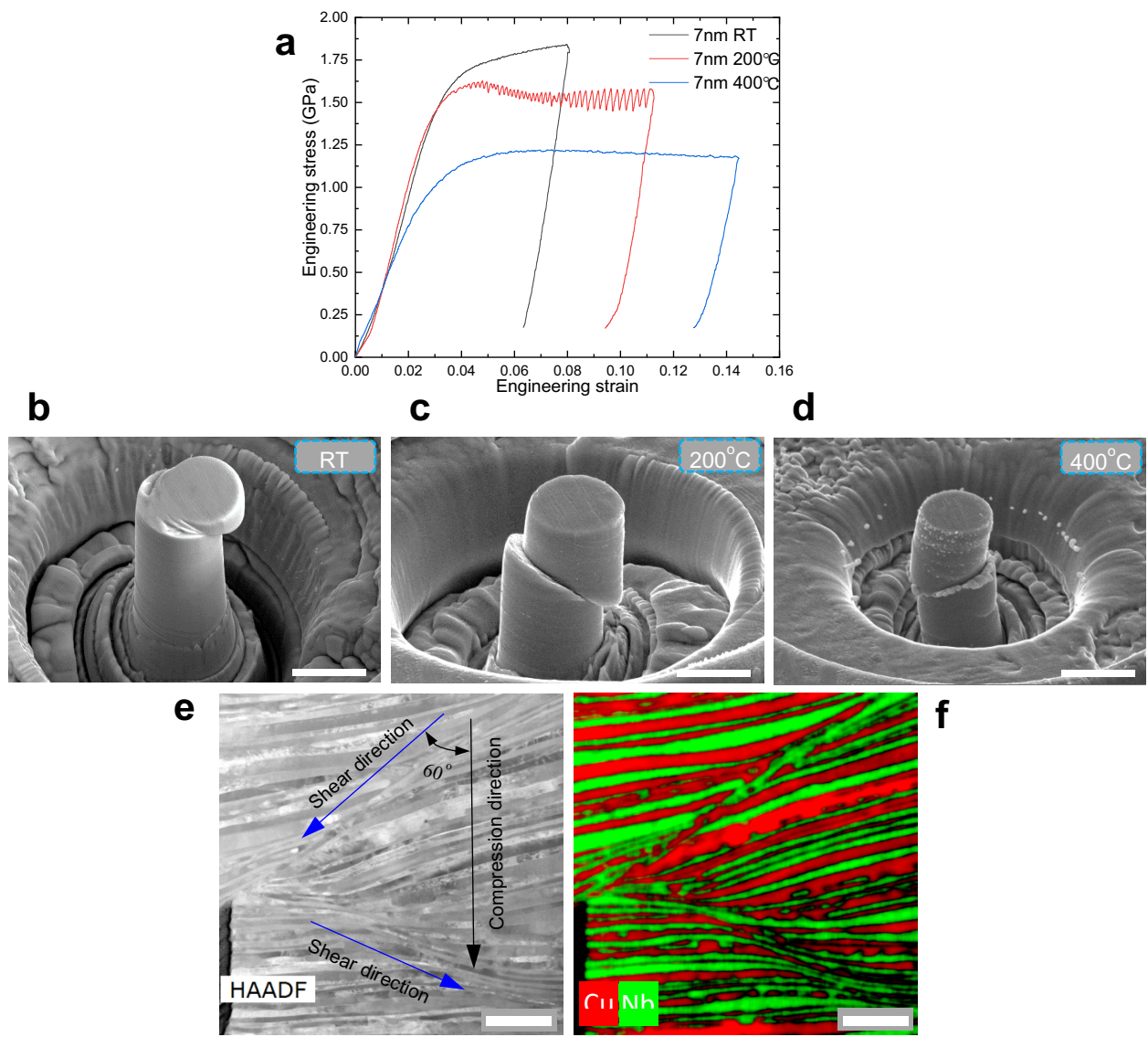

**Fig. 6.** (a) Engineering stress-strain curves of Cu/Nb MNL with 7 nm layer thickness obtained from micropillar compression tests at constant strain rate of $10^{-3}$/s at room temperature (RT), 200 °C and 400 °C. Serrations only were observed in the micropillar tested at 200 °C. (b) SEM of the Cu/Nb MNL micropillar of 7 nm layer thickness deformed at RT. (c) *Idem* at 200 °C. (d) *Idem* at 400 °C. Cu diffusion on the micropillar surface is visible, particularly in highly-stressed areas. (e) HAADF-STEM image of shear bands in the micropillar deformed at 200 °C. (f) EDS map showing the distribution of Cu and Nb in (e). The scale bars are 5 μm in (b), (c) and (d), and 200 nm in (e) and (f).

In order to ascertain the origin of the shear bands generated during micropillar compression at 200 °C, a thin foil parallel to the micropillar containing the shear band was lifted out using focused-ion beam (FIB) and analyzed in the transmission electron microscope (Fig. 1). The high angle annular dark field (HAADF) scanning-transmission electron microscope (STEM) image of the region containing the shear band is depicted in Fig. 6e. The chemical composition of each layer (either Cu or Nb) within this region can be ascertained from the energy dispersive spectroscopy (EDS) map in Fig. 6f. The highly deformed shear zones broke the continuity of individual Cu or Nb layers and were mainly formed along the Nb layers. Similar images and EDS maps in other shear bands in the Cu/Nb MNL with 7 nm average layer thickness deformed at 200 °C can be found in Fig. 7.

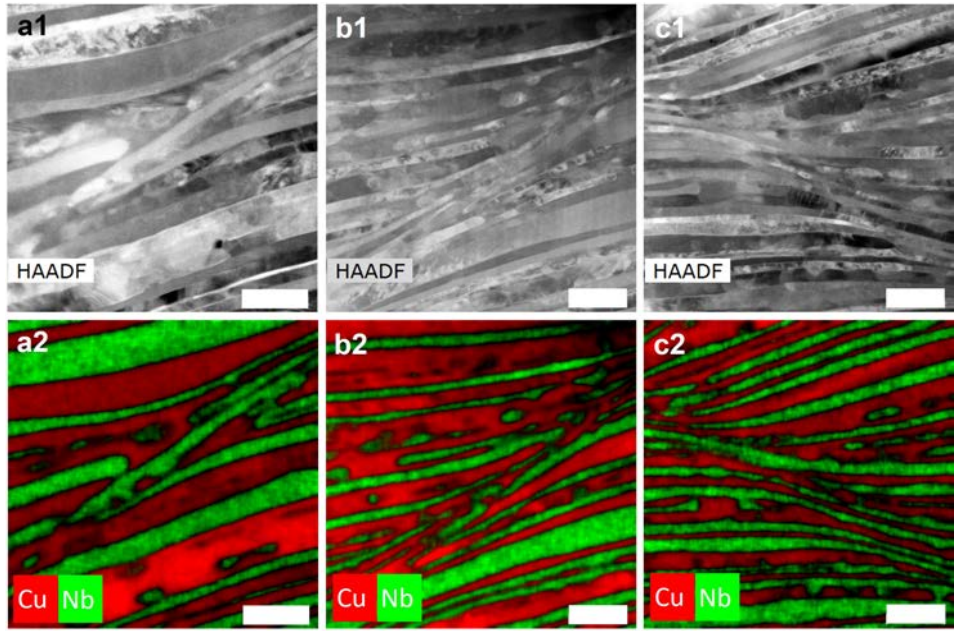

**Fig. 7.** TEM bright-field image of the shear zones in Cu/Nb MNL with 7 nm layer thickness deformed at 200 ºC. (a1), (b1) and (c1) HAADF STEM images of the shear zones that disrupt the continuity of the MNL layers. (a2), (b2) and (c2) Energy dispersive spectroscopy map showing the elemental distribution of Cu and Nb in (a1), (b1) and (c1), respectively. The scale bars in (a1) and (a2) are 80 nm, and 100 nm for (b1), (b2), (c1) and (c2).

Needle samples were manufactured using FIB around the shear bands of deformed micropillars for atom-probe tomography (APT) examination, following the procedure depicted in Fig. 2. They were extracted from the Cu/Nb MNL with 7 nm average layer thickness deformed at RT, 200 ºC and 400 ºC. The spatial local distribution of Cu, Nb and O atoms in three APT needles is shown in Figs. 8a, 8b and 8c for the specimens deformed at each temperature, respectively. They were the only major elements found in the needle and their spatial distribution along the blue cylinders perpendicular to the Cu/Nb interfaces are plotted in Figs. 8d, 8e and 8f. In all cases, and regardless of the temperature, the O content in the Cu layers was negligible. On the contrary, O was always found within the Nb layers and the O content increased with the test temperature: from 2% in the micropillar deformed at RT to 2.5% after deformation at 200 ºC and up to 10% in the micropillar deformed at 400 ºC. Moreover, the Cu/Nb interfaces were very sharp in the specimens deformed at RT and 200 ºC but became wider (due to diffusion of Cu and Nb in the samples deformed at 400 ºC. It should be noted that O atoms were always found in the Nb layers while the O content in Cu was negligible and the concentration of O atoms was very uniform in the Nb nanolayer at 200 ºC (~ 2.5 at.%). The concentration of O in Nb increased with the distance to the Cu/Nb interface in the specimens deformed at 400 ºC and other APT analysis (depicted in Fig. 9) showed that a significant fraction of the O in this case has led to the formation of $NbO_2$.

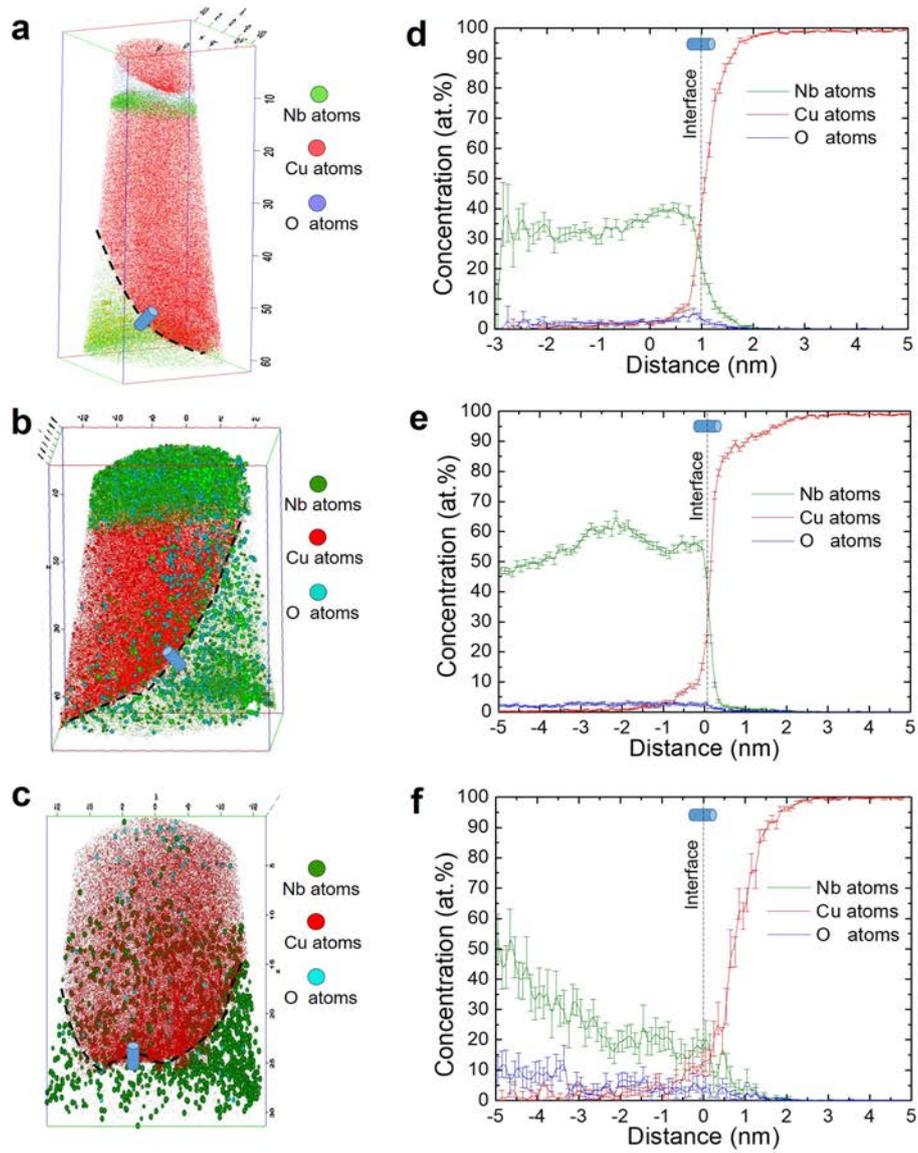

**Fig. 8.** (a) Spatial distribution of Cu, Nb and O in an APT needle extracted from a Cu/Nb micropillar with 7 nm layer thickness deformed at RT. (b) *Idem* as (a) in a micropillar deformed at 200 ºC. (c) *Idem* as (a) in a micropillar deformed at 400 ºC. (d) Concentration of Cu, Nb and O along the blue cylinder in (a) perpendicular to the Cu/Nb interface. (e) *Idem* as (d) along the blue cylinder perpendicular to the interface in (b). (f) *Idem* as (d) along the blue cylinder perpendicular to the interface in (c). The Cu/Nb interface in each needle is marked by a dashed black line.

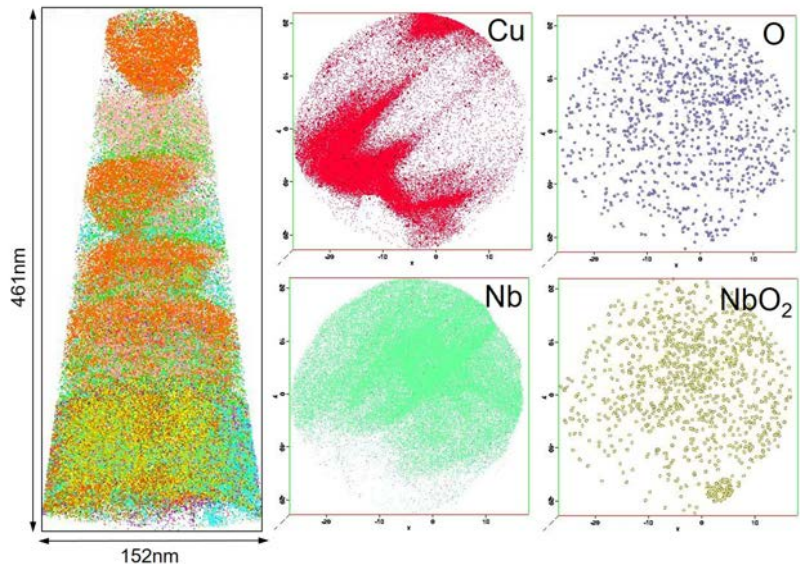

**Fig. 9.** APT mapping of one Cu/Nb nanolaminate micropillar (with 7 nm average layer thickness) deformed at 400 ºC, showing Cu, Nb, O and $NbO_2$ distribution after deformation. Considerable Nb atoms are transformed into $NbO_2$.

## 4. Discussion

Constant strain rate tests as well as strain-rate jump tests in micropillars of Cu/Nb MNL showed compelling evidence of DSA when deformation took place at 200 ºC in MNLs with average layer thicknesses of 7, 16 and 34 nm. They include the presence of serrations in the stress-strain curves, the localization of the plastic deformation along shear bands when the applied strain was small (< 6%), and the reduction of the strain rate sensitivity to 0. Nevertheless, DSA was not observed either for the Cu/Nb MNL deformed at RT and 400 ºC or for the Cu/Nb MNL with 63 nm layer thickness deformed at RT, 200 ºC and 400 ºC. Moreover, APT analysis showed the presence of O (around 2- 2.5%) in the Nb layers of the MNL that have been tested at RT and 200 ºC and while more O was found in the Nb in the samples deformed at 400 ºC, O was never found in the Cu layers.

Electrical resistivity studies by Köthe [27] provided evidence of the formation of interstitial impurity atmospheres of O around dislocations in Nb in the temperature range 140 ºC to 220 ºC in samples containing 160 ppm of O and the number of O atoms bound to dislocations varied linearly with $1/T$, being $T$ the absolute temperature. This information is agreement with the experimental evidence of DSA in Nb in the temperature range ≈ 150 – 300 ºC due to the formation of Cottrell atmospheres of O atoms around the dislocations [21-23]. Thus, the presence of O in the Nb layers was responsible for the development of DSA in the Cu/Nb nanolaminates with 7, 16 and 34 nm layer thickness deformed at 200 ºC. The interstitial O atoms diffused to the dislocations in Nb and hindered dislocation transmission at the interface. This was the dominant deformation mechanisms at RT for the MNL with 7 and 16 nm layer thickness and, very likely, it was also the dominant one of the MNL with 34 nm layer thickness at 200 ºC because the maximum compressive stress in all cases was very similar (around 1.5-1.6 GPa). Nevertheless, DSA was not observed in the Cu/Nb MNL with 64 nm layer thickness deformed at 200 ºC and the mechanical response at RT and 200 ºC was very similar. According to the activation volume and the strength level (0.8-1.0 GPa), deformation of this MNL was controlled by confined layer slip in both Cu and Nb layers. Nevertheless, recent *in-situ* TEM mechanical tests of Cu/Nb MNL with 63 nm layer thickness manufactured by ARB showed that - although confined layer slip in both Cu and Nb took place - most of the deformation tended to be localized in the thicker Cu layers when the MNL was deformed perpendicular to the layers [12]. Thus, deformation of the softer Cu layers controlled the mechanical response, rather than the deformation of the Nb layers, and DSA was not observed. This was not the case, however, in the Cu/Nb MNLs with smaller layer thickness (7, 16 and 34 nm) where deformation of the Nb layers by either dislocation transmission across the interfaces or confined layer slip was necessary, leading to the appearance of DSA.

## 5. Conclusions

In summary, evidence of DSA has been reported for the first in Cu/Nb MNLs tested at 200 ºC due to the diffusion of O into the Nb layers. DSA occurred whether the main deformation mechanism of the Cu/Nb MNL was either dislocation transmission across the interface or confined layer slip and took place around 200 ºC. DSA was not found either at RT because the diffusivity of the O atoms towards the mobile dislocations was too low, or at 400 ºC because the deformation was mainly localized in the Cu layers following a power-law creep regime. Deformation of the Cu/Nb MNL at 400 ºC was mainly controlled by the stress-assisted diffusion of Cu. This process was enhanced by the oxidation of Nb layers to form $NbO_2$, which limited more the contribution of Nb to the overall plastic deformation. Thus, DSA was not found in the Cu/Nb MNL deformed at 400 ºC. Finally, DSA was neither observed in the Cu/Nb MNL with 64 nm average layer thickness deformed at 200 ºC because deformation was mainly localized in the thicker Cu layers within the MNL. DSA led to a very high strength retention (> 1.50 GPa) of the Cu/Nb MNLs with 7, 16 and 34 nm layer thickness at 200 ºC, very close to the ones reported at RT, although it also promoted early failure due to shear localization as a result in the reduction of the strain rate sensitivity. These novel findings show how the mechanical response of ARB MNL can be tailored through the application of DSA.

## Declaration of interest

The authors declare that they have no known competing financial interests or personal relationships that could have appeared to influence the work reported in this paper.

## Acknowledgements

This work was supported by the European Research Council (ERC) under the European Union's Horizon 2020 research and innovation programme (Advanced Grant VIRMETAL, grant agreement No. 669141). Dr. Z. Liu acknowledges European Union for the Marie Sklodowska-Curie Individual Fellowship with the project MINIMAL (grant agreement No. 749192), and thanks the support from Innovation Driven Program of Central South University (Grant No. 2019CX006). All authors thank Dr. N. A. Mara and Dr. I. J. Beyerlein for providing the Cu/Nb nanolaminates, to Dr. M. Castillo-Rodríguez for FIB and TEM operation assistance and to Dr. O. Renk for useful discussions about dynamic strain aging in Nb.

## Data availability

The original data of this study can be obtained by request to the corresponding author.

# Supplementary materials

Zhilin Liu, J. Snel, T. Boll, J.-Y. Wang, M. A. Monclús, J. M. Molina-Aldareguía, J. LLorca. High temperature strength retention of Cu/Nb nanolaminates through dynamic strain ageing.

**Video S1.** In-situ compression test of the Cu/Nb MNL micropillar with 7 nm average layer thickness deformed at room temperature.

**Video S2.** In-situ compression test of the Cu/Nb MNL micropillar with 7 nm average layer thickness deformed at 200 ºC, showing evidence of dynamic strain ageing.

**Video S3.** In-situ compression test of the Cu/Nb MNL micropillar with 7 nm average layer thickness deformed at 400 ºC.